\documentclass[12pt,oneside]{article}

\usepackage{cite}
\usepackage{acronym}
\usepackage{enumerate}  
\usepackage{a4wide}
\usepackage[left=2.5cm, right=2.5cm, top=2.5cm, bottom=2.5cm]{geometry}
\usepackage{fancyhdr}
\usepackage{graphicx}
\usepackage{palatino}
\usepackage{blindtext}
\usepackage{multirow}
\usepackage{amsmath}
\usepackage{amssymb}
\usepackage{mathrsfs}
\usepackage{bbm}
\usepackage{ragged2e}

\usepackage[T1]{fontenc}
\usepackage[utf8]{inputenc}
\usepackage[bookmarks]{hyperref}
\usepackage[justification=centering]{caption}
\usepackage{csquotes}
\usepackage[ngerman]{cleveref}

\usepackage{tikz}
\usetikzlibrary{calc}
\usetikzlibrary{decorations.pathmorphing}
\tikzset{snake it/.style={decorate, decoration=snake}}

\newtheorem{theorem}{Theorem}

\newtheorem{proposition}[theorem]{Proposition}
\newtheorem{corollary}[theorem]{Corollary}

\newcommand{\qedsymbol}{\hfill\square}

\title{Kodama-like Vector Fields in \\ 
Axisymmetric Spacetimes}
\author{{Philipp Dorau}${}^a$\footnote{\href{mailto:philipp.dorau@uni-leipzig.de}{philipp.dorau@uni-leipzig.de}} \ \, and \ \, {Rainer Verch}${}^{a,b}$\footnote{\href{mailto:rainer.verch@uni-leipzig.de}{rainer.verch@uni-leipzig.de}}\\[6pt]
\small{${}^a$\, Institut f\"ur Theoretische Physik, Universit\"at Leipzig, 04103 Leipzig, Germany}\\
\small{${}^b$\, CY Advanced Studies, Cergy Paris 
Universit\'e, 95000 Neuville-sur-Oise, France}
}
\date{June 18, 2024}

\begin{document}

\maketitle

\begin{abstract}
We extend the concept of the Kodama symmetry, a quasi-local time translation symmetry for dynamical spherically symmetric spacetimes, to a specific  class of  dynamical axisymmetric spacetimes, namely the families of Kerr-Vaidya and Kerr-Vaidya-de Sitter spacetimes. We study some geometrical properties of the asymptotically flat Kerr-Vaidya metric, such as the Brown-York mass and the Einstein tensor. Furthermore, we propose a generalization of the Kerr-Vaidya metric to an asymptotic de Sitter background. We show that for these classes of dynamical axisymmetric black hole spacetimes, there exists a timelike vector field that exhibits similar properties to the Kodama vector field in spherical symmetry. This includes the construction of a covariantly conserved current and a corresponding locally conserved charge, which in the Kerr-Vaidya case converges to the Brown-York mass in the asymptotically flat region.
\end{abstract}

\newpage
\section{Introduction}\label{Introduction}
It is widely believed that physical black holes are dynamical, e.g. they accrete classical matter \cite{NQ:2005acc}, or undergo semi-classical evaporation processes via Hawking radiation \cite{Hawking:1975pc, Unruh:1976evap, Page:2005bhtd, KHZ:2021dc, JV:2023evap}, although the latter remains a hypothetical theoretical scenario for the time being. Therefore, the mathematical aspects of dynamical, non-stationary black hole spacetimes are an active field of research, aiming at a physically more accurate description of black hole evolution. Various authors have been able to derive very general statements about dynamical black holes, e.g. the general laws of black hole dynamics \cite{Hayward:1994bhd}. \\

In this work, we focus on what is often referred to as the \emph{Kodama miracle}: In a general non-stationary spherically symmetric spacetime, there exists a timelike vector field which gives rise to a covariantly conserved current when contracted with the Einstein tensor of the metric \cite{Kodama:1980og}, called the Kodama vector field. Hence, the Kodama vector field generates a time translation symmetry and is the non-stationary analogue to a timelike Killing vector field. Furthermore, it is argued that the Kodama flow provides a geometrically preferred direction of time in spherically symmetric dynamical spacetimes \cite{AV:2010kt}. \\

Since its discovery, the Kodama vector field is widely used in the study of dynamical black holes, e.g. as a time evolution vector field for field theories in black hole spacetimes, see \cite{Racz:2006kod}. Especially in the field of semi-classical gravity and black hole evaporation, the Kodama vector has proven to be very useful. For instance, recent works use the existence and the properties of the Kodama vector field to compute the relative entropy between two coherent states of a quantum field theory in spherically symmetric dynamical spacetimes, see \cite{DAngelo:2021bhre,KPV:2021ea}. \\

In the main results of this paper, we provide a first generalization of the Kodama symmetry to a simple class of axisymmetric dynamical spacetimes. More specifically, we construct a Kodama-like vector field in the Kerr-Vaidya family of spacetimes, see \cite{DT:2020kv,Dahal:2021kvh,DMST:2023KVii}. This timelike vector field qualitatively possesses the same properties as the Kodama vector field in spherical symmetry. Most importantly, its associated current is covariantly conserved. For this Kodama-like current, we derive a relation between the corresponding Noether charge and the Brown-York mass of the spacetime. Additionally, we use these results to compute an expression for the angular momentum in Kerr-Vaidya spacetime. \\

Ultimately, we extend our results from the asymptotically flat case to a cosmological background, i.e. Kerr-Vaidya-de Sitter spacetime. We make a (as far as we are able to tell, original) proposal for spacetimes of Kerr-Vaidya-de Sitter type, and we deduce that a Kodama-like vector field also exists in Kerr-Vaidya-type black holes with an asymptotic de Sitter background. We find a relation between the Kodama-like current and its associated charges, which is qualitatively similar to the asymptotically flat case. However, some properties of the charges, such as the large-$r$ behaviour, do not necessarily carry over to the asymptotically de Sitter case.

\section{Kerr-Vaidya Spacetime}\label{KV}
In this section, we want to provide an overview of a particularly interesting case of an axisymmetric dynamical black hole spacetime. More specifically, we shortly explain the derivation and the properties of the Kerr-Vaidya metric, a dynamical version of the Kerr metric where the constant mass $M$ is replaced by a function $M(v)$ of advanced time $v$. \\

Let us recall the Newman-Janis algorithm \cite{NJ:1965nj} to derive the rotating Kerr metric from the static Schwarzschild metric via a complex coordinate transformation, using the New\-man-Penrose tetrad formalism \cite{NP:1962f}. In an analogous manner, it is then possible to derive a rotating dynamical metric from the simpler Vaidya metric, see \cite{Vaidya:1953bh}, which is the non-stationary generalization of the Schwarzschild metric \cite{GHJ:1979kv, HJ:1982nj, DT:2020kv}. \\

Remarkably, the same result for a rotating Vaidya-type metric is obtained by the following procedure. Firstly, we take the stationary Kerr metric in Boyer-Lindquist coordinates $(t,r,\vartheta,\varphi) \in \mathbb{R}\times \mathbb{R}_+ \times(0,\pi)\times(0,2\pi)$ \cite{BL:1967coord}
\begin{equation}
ds^2 = -\left(1-\frac{2Mr}{\rho^2}\right) dt^2 - \frac{4Mar \sin^2 \vartheta}{\rho^2} dt d\varphi + \frac{\rho^2}{\Delta} dr^2 + \rho^2 d\vartheta^2 + \frac{\Gamma^2 - \Delta a^2 \sin^2 \vartheta}{\rho^2}\sin^2\vartheta d\varphi^2,
\end{equation}

\noindent which are singular at $\Delta = 0$ and $\rho^2 = 0$, where $\rho^2, \Delta, \Gamma$ are given by the expressions
\begin{align}
\rho^2 &= r^2 + a^2 \cos^2 \vartheta , \\
\Delta &= r^2 - 2Mr + a^2 \label{Delta} , \\
\Gamma &= r^2 + a^2,
\end{align}

\noindent $M>0$ describes the mass of the black hole, and $a\in\mathbb{R}$ with $\vert a \vert \leq M$ denotes the rotation parameter. We then transform to ingoing Eddington-Finkelstein-type coordinates $(v,r,\vartheta,\psi) \in \mathbb{R}\times\mathbb{R}_+ \times(0,\pi)\times(0,2\pi)$, which are convenient for the description of the black hole region since they are regular at the future event horizon, see \cite{Poisson:2004tool,DMST:2023KVii}. Explicitly, this transformation is given by
\begin{align}
dv &= dt + \frac{\Gamma}{\Delta}dr , \\
d\psi &= d\varphi + \frac{a}{\Delta}dr,
\end{align}

\noindent which leads to the metric \cite{Poisson:2004tool}
\begin{align}
ds^2 = &-\left(1-\frac{2Mr}{\rho^2}\right) dv^2 + 2dvdr - \frac{4Mar \sin^2 \vartheta}{\rho^2} dv d\psi - 2a\sin^2\vartheta dr d\psi \nonumber \\
&+ \rho^2 d\vartheta^2 + \frac{\Gamma^2 - \Delta a^2 \sin^2 \vartheta}{\rho^2}\sin^2\vartheta d\psi^2 .
\end{align}

\noindent Finally, we obtain the Kerr-Vaidya metric by assigning a dynamic behaviour to the mass, i.e. we replace the constant parameter $M$ by a suitable function $M(v)$ of the advanced time coordinate $v$, in order to obtain \cite{DT:2020kv, DMT:2022scbh, MMT:2022scbh}

\begin{align}
ds^2 = &-\left(1-\frac{2M(v)r}{\rho^2}\right) dv^2 + 2dvdr - \frac{4M(v)ar \sin^2 \vartheta}{\rho^2} dv d\psi - 2a\sin^2\vartheta dr d\psi \nonumber \\
&+ \rho^2 d\vartheta^2 + \frac{\Gamma^2 - \Delta a^2 \sin^2 \vartheta}{\rho^2}\sin^2\vartheta d\psi^2.
\label{KerrVaidyaMetric}
\end{align}

\noindent Note that the replacement $M\rightarrow M(v)$ is also implied for the function $\Delta$ given in \eqref{Delta}. \\

Moreover, we want to summarize some important geometric properties of the Kerr-Vaidya metric \eqref{KerrVaidyaMetric}. In contrast to the Kerr metric, the Kerr-Vaidya metric is non-stationary, i.e. it does not possess a timelike Killing vector field.\ However, the Kerr-Vaidya metric remains to be axisymmetric, i.e. it possesses the spacelike Killing vector field
\begin{equation}
\label{AxialKillingField}
\phi^a = \left( \frac{\partial}{\partial \psi} \right)^a.
\end{equation}

\noindent Additionally, as for the stationary case, the square root of the (negative) determinant of the Kerr-Vaidya metric is given by
\begin{equation}
\sqrt{-g} = \rho^2 \sin\vartheta.
\end{equation}

\begin{theorem}\label{Theorem_BY}
The Brown-York quasi-local mass of Kerr-Vaidya spacetime is given by the $v$-dependent dynamical mass, i.e.
\begin{equation}
M_\textrm{BY} = M(v).
\end{equation}

\end{theorem}

\noindent The proof of this statement is, together with the definition of $M_\textrm{BY}$, given in Appendix \ref{Proof_BY}. \\

Proceeding to general relativistic curvature quantities, we find that the Kerr-Vaidya metric has a vanishing Ricci scalar, i.e. $R=0$. More generally, however, the corresponding Einstein tensor $G_{ab}$ does not vanish. In the ingoing coordinate system $(v,r,\vartheta,\psi)$, it is of the schematic form \cite{DT:2020kv}
\begin{equation}
\label{KVET}
(G_{\mu\nu}) = \begin{pmatrix}
G_{vv} & 0 & G_{v\vartheta} & G_{v\psi} \\
0 & 0 & 0 & 0 \\
G_{\vartheta v} & 0 & 0 & G_{\vartheta\psi} \\
G_{\psi v} & 0 & G_{\psi \vartheta} & G_{\psi\psi}
\end{pmatrix}  ,
\end{equation}

\noindent with the $vv$-component explicitly given by \cite{DT:2020kv, Mathematica, xAct, xPerm, xTensor, xCoba}
\begin{equation}
G_{vv} = \frac{1}{\rho^6} \left( 2M'(v) \left( a^4 (\cos^4 \vartheta - \cos^2 \vartheta ) + \Gamma r^2 \right) - M''(v) r \rho^2 a^2 \sin^2 \vartheta \right).
\end{equation}

\noindent where $M'(v)$ denotes the first derivative of $M(v)$ with respect to advanced time $v$, and $M''(v)$ denotes the second derivative, respectively. This means that the Kerr-Vaidya metric only solves the Einstein equations for the non-vacuum stress energy tensor corresponding to expression \eqref{KVET}, which is of type III in the Hawking-Ellis classification \cite{DT:2020kv}, and thus violates all of the standard energy conditions \cite{HE:1973st, MV:2017ec, MV:2018iii}. Hence, we conclude that the greatest relevance of the Kerr-Vaidya spacetime is in the context of quantum effects, e.g. black hole evaporation, instead of purely classical considerations. Additionally, recent works suggest that this type of exotic matter enables a finite observable formation time of the Kerr-Vaidya black hole \cite{DMST:2023KVii}. Alternatively, it has been shown that type III stress energy tensors are realized more easily in classical scalar-tensor theories of gravity, see \cite{BFVG:2023st}, which further adds to the applicability of the model, aside from quantum theory.

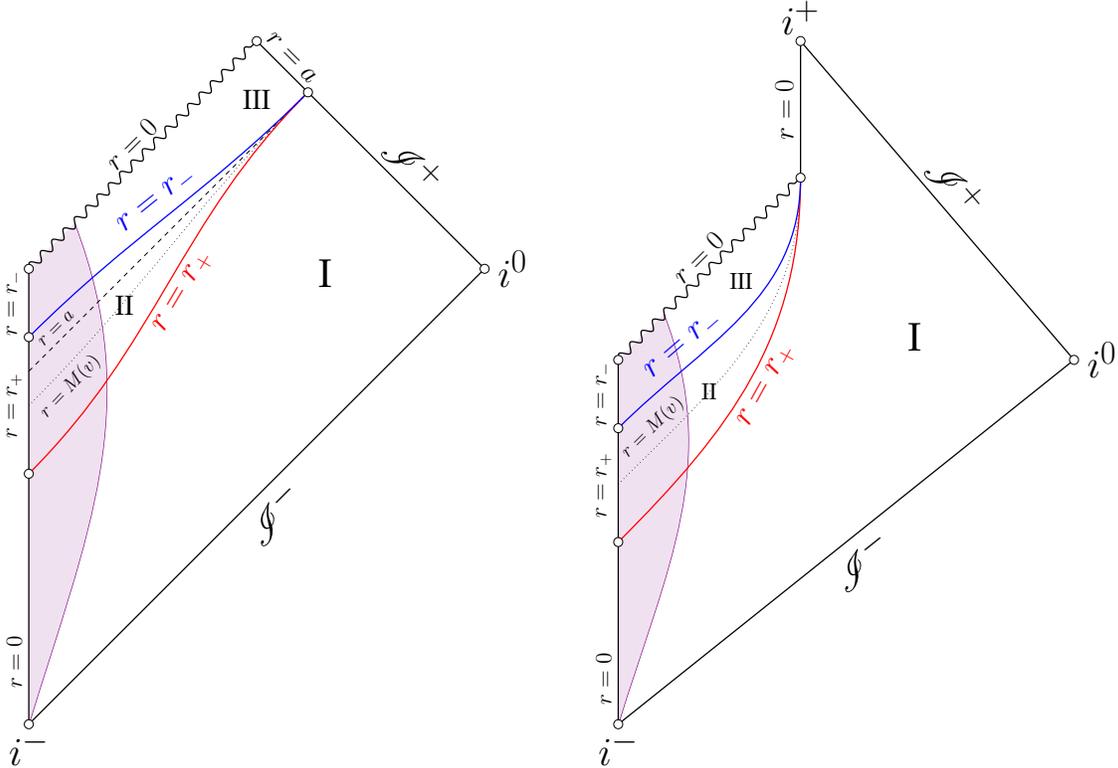
\begin{figure}[h!]
\begin{minipage}[t]{0.5\linewidth}
\centering
\resizebox{0.87\textwidth}{!}{
\begin{tikzpicture}
\coordinate (i-) at (0,0);
\coordinate (r+) at (0,5.5);
\coordinate (rM) at (0,7);
\coordinate (ra) at (0,7.75);
\coordinate (r-) at (0,8.5);
\coordinate (lefttop) at (0,10);
\coordinate (pcollapse) at (1,11);
\coordinate (Iright) at (10,10);
\coordinate (top) at (5,15);
\coordinate (exfin) at (6.125,13.875);
\draw[thick, violet] (i-) to[out=73,in=-69] (pcollapse);
\fill[violet!12, opacity=0.3] (i-) to[out=73, in=-69] (pcollapse) -- (lefttop) -- cycle;
\draw[thick] (i-) -- node[near start, above, sloped, font=\large]{$r=0$} (r+);
\draw[thick] (r+) -- node[midway, above, sloped, font=\large]{$r=r_+$} (r-);
\draw[thick] (r-) -- node[midway, above, sloped, font=\large]{$r=r_-$} (lefttop);
\draw[thick, snake it] (lefttop) -- node[midway, above, sloped, font=\Large]{$r=0$} (top);
\draw[thick] (top) -- node[midway, above, sloped, font=\Large]{$r=a$} (exfin);
\draw[thick] (exfin) -- node[midway, above, sloped, font=\huge]{$\mathscr{I}^+$} (Iright);
\draw[thick] (Iright) -- node[midway, below, sloped, font=\huge]{$\mathscr{I}^-$} (i-);
\draw[dashed] (ra) -- node[pos=0.127, above, sloped, font=\normalsize]{$r=a$} (exfin);
\draw[dotted] (rM) to[out=45,in=-135] node[pos=0.09, below, sloped, font=\normalsize]{$r=M(v)$} (exfin);
\draw[thick, red] (r+) to[out=45,in=-135] node[midway, below, sloped, font=\LARGE]{$r=r_+$} (exfin);
\draw[thick, blue] (r-) to[out=45,in=-135] node[midway, above, sloped, font=\LARGE]{$r=r_-$}(exfin); 
\node[below] at (i-) {\huge $i^-$};
\node[right] at (Iright) {\huge $\, i^0$};
\node[font=\huge] at (6.5,9.9) {I};
\node[font=\Large] at (2.1,9.2) {II};
\node[font=\Large] at (5,13.73) {III};
\draw[black,fill=white] (0,0) circle (.5ex);
\draw[black,fill=white] (r+) circle (.5ex);
\draw[black,fill=white] (r-) circle (.5ex);
\draw[black,fill=white] (lefttop) circle (.5ex);
\draw[black,fill=white] (top) circle (.5ex);
\draw[black,fill=white] (exfin) circle (.5ex);
\draw[black,fill=white] (Iright) circle (.5ex); 
\end{tikzpicture}
}
\end{minipage}
\hfill
\begin{minipage}[t]{0.5\linewidth}
\resizebox{0.87\textwidth}{!}{
\begin{tikzpicture}
\coordinate (i-) at (0,0);
\coordinate (r+) at (0,4);
\coordinate (rM) at (0,5.27);
\coordinate (r-) at (0,6.5);
\coordinate (lefttop) at (0,8);
\coordinate (pcollapse) at (1,9);
\coordinate (Iright) at (10,8);
\coordinate (topright) at (4,12);
\coordinate (top) at (4,15);
\draw[thick, violet] (i-) to[out=73,in=-69] (pcollapse);
\fill[violet!12, opacity=0.3] (i-) to[out=73, in=-69] (pcollapse) -- (lefttop) -- cycle;  
\draw[thick] (i-) -- node[near start, above, sloped, font=\large]{$r=0$} (r+);
\draw[thick] (r+) -- node[midway, above, sloped, font=\large]{$r=r_+$} (r-);
\draw[thick] (r-) -- node[midway, above, sloped, font=\large]{$r=r_-$} (lefttop);
\draw[thick, snake it] (lefttop) -- node[midway, above, sloped, font=\Large]{$r=0$} (topright);
\draw[thick] (topright) -- node[midway, above, sloped, font=\Large]{$r=0$} (top);
\draw[thick] (top) -- node[midway, above, sloped, font=\huge]{$\mathscr{I}^+$} (Iright);
\draw[thick] (Iright) -- node[midway, below, sloped, font=\huge]{$\mathscr{I}^-$} (i-);
\draw[dotted] (rM) to[out=45,in=-90] node[pos=0.16, above, sloped, font=\normalsize]{$r=M(v)$} (topright);
\draw[thick, red] (r+) to[out=45,in=-90] node[midway, below, sloped, font=\LARGE]{$r=r_+$} (topright);
\draw[thick, blue] (r-) to[out=45,in=-90] node[pos=0.3, above, sloped, font=\LARGE]{$r=r_-$}(topright); 
\node[right] at (Iright) {\huge $\,i^0$};
\node[above] at (top) {\huge $i^+$};
\node[below] at (i-) {\huge $i^-$};
\node[font=\huge] at (6.5,8.5) {I};
\node[font=\large] at (2.7,9.73) {III};
\node[font=\large] at (2,7.3) {II};
\draw[black,fill=white] (0,0) circle (.5ex);
\draw[black,fill=white] (r+) circle (.5ex);
\draw[black,fill=white] (r-) circle (.5ex);
\draw[black,fill=white] (lefttop) circle (.5ex);
\draw[black,fill=white] (top) circle (.5ex);
\draw[black,fill=white] (topright) circle (.5ex);
\draw[black,fill=white] (Iright) circle (.5ex); 
\end{tikzpicture}
}
\end{minipage}
\caption{\justifying Schematic Carter-Penrose diagrams for asymptotically flat sub-extremal Kerr-Vaidya-type spacetimes, showing rotating black holes which form by gravitational collapse and subsequently evaporate via Hawking radiation. The left diagram shows an asymptotically decreasing evaporation process towards extremality, i.e. $M(v)\rightarrow a$ for $v \rightarrow \infty$. This scenario requires that the Hawking temperature converges to zero for $M(v)\rightarrow a$. The diagram on the right depicts the complete evaporation of the black hole, see \cite{SBD:2023pd}, for which a dynamically decreasing rotation parameter $a(v)$ is necessary.}
\end{figure}

The trapping horizons of stationary Kerr black holes, see \cite{Faraoni:2015horizons}, coincide with the event horizons, which are given by the solutions
\begin{equation}
\label{KerrOuterInnerHorizon}
r_\pm = M \pm \sqrt{M^2 - a^2}
\end{equation}

\noindent of $\Delta = 0$. However, the trapping of light is substantially more complicated in the dynamical case, since for $M'(v) \neq 0$ the trapping horizons of the black hole do not coincide with $r_\pm$ anymore. For more detailed illustrations of this issue and an approximate expression for an apparent horizon, we refer to \cite{ST:2015pp, DT:2020kv, Dahal:2021kvh, DMST:2023KVii}.

\section{Kodama Symmetry in Spherically Symmetric Spacetimes}\label{SectionKodama}
By definition, non-stationary spacetimes do not possess a timelike Killing vector field. Nonetheless, in spherically symmetric spacetimes, there exists a timelike vector field $k^a$, called the Kodama vector, which hints at a different type of time translation symmetry, even for dynamical spacetimes \cite{Kodama:1980og}. \\

More specifically, let us consider a $(3+1)$-dimensional spherically symmetric spacetime with a metric of the form
\begin{equation}
ds^2 = g_{ij} dx^i dx^j + R^2 (x^1, x^2) \left(d\vartheta^2 + \sin^2 \vartheta d\varphi^2 \right),
\end{equation}

\noindent for coordinates $(x^1, x^2, \vartheta, \varphi)\in \mathbb{R}^2 \times \mathbb{S}^2$. Here $g_{ij}$ denotes a $2$-dimensional non-singular Lorentzian metric, and $R$ is a smooth function corresponding to the radius. On this general spherically symmetric spacetime, the Kodama vector is given by \cite{Kodama:1980og, Racz:2006kod, AV:2010kt}
\begin{equation}
\label{ClassicalDefinitionKodama}
k^a = \epsilon^{ab} \nabla_b R,
\end{equation}

\noindent where $\nabla$ denotes the covariant derivative corresponding to the spacetime metric, and $\epsilon_{ab}$ denotes the canonically embedded volume form of the $2$-dimensional metric $g_{ij}$, see \cite{Wald:1984gr}. \\

We want to point out three particularly important properties of the Kodama vector field. First of all, $k^a$ is divergence-free, i.e.
\begin{equation}
\nabla_a k^a = 0.
\end{equation}

\noindent Furthermore, it lies tangent to surfaces of constant $R$, i.e. it fulfills the relation
\begin{equation}
k^a \nabla_a R = 0.
\end{equation}

\noindent However, perhaps the most important property of $k^a$ is that, together with the Einstein tensor $G_{ab}$ of the spacetime metric, it gives rise to a current
\begin{equation}
j^a = G^{ab} k_b,
\end{equation}

\noindent which is covariantly conserved \cite{Racz:2006kod, AV:2010kt}, i.e.
\begin{equation}
\nabla_a j^a = G^{ab} \nabla_a k_b = 0.
\end{equation}

\noindent In other words, this means that even in the absence of a timelike Killing vector field, the Kodama vector $k^a$ provides a locally conserved flux of energy $j^a$ \cite{Racz:2006kod}. \\

Hence, the existence of the Kodama vector field in spherical symmetry yields useful applications in the study of dynamical black holes. These include a definition of a Kodama surface gravity which, in turn, can be associated to thermodynamic quantities, such as temperature and entropy of a black hole horizon \cite{Hayward:1998bhtd}. Additionally, the conserved current $j^a$, related to the Kodama vector $k^a$, can be viewed as a Noether current, with its corresponding charge being the Misner-Sharp-Hernandez mass, i.e. the gravitational energy of the spherically symmetric black hole \cite{Hayward:1996sphsym}. \\

Another prominent feature of the Kodama vector $k^a$ is that it induces a naturally preferred time direction. In other words, this means that the choice of a time coordinate for the spherically symmetric dynamical spacetime along the direction of the Kodama flow is geometrically special \cite{AV:2010kt}. \\

While the Kodama formalism is well-established in spherical symmetry, it is not yet clear whether general Kodama-like vector fields also exist outside of spherical symmetry. A notable result for a Kodama vector field in a general $(2+1)$-dimensional axisymmetric spacetime has recently been obtained by Kinoshita, see \cite{Kinoshita:2021axis}. However, one might argue that axisymmetry in $(2+1)$ dimensions is the equivalent of spherical symmetry in $(3+1)$ dimensions. Therefore, our goal is to find Kodama-like vector fields for specific cases of $(3+1)$-dimensional axisymmetric spacetimes, in particular, the Kerr-Vaidya geometry discussed in Section \ref{KV}.

\section{Kodama-like Symmetry in Kerr-Vaidya Spacetime}\label{KodamaLike}
First of all, we aim to find a timelike vector field $K^a$ in Kerr-Vaidya spacetime that fulfills the same properties as $k^a$ in the spherically symmetric setting, including a covariantly conserved current $J^a = G^{ab}K_b$ and a corresponding charge $Q_K$.

\begin{proposition}\label{KodamaKV}
The vector field
\begin{equation}
\label{KodamaResult}
K^a = \left( \frac{\partial}{\partial v} \right)^a
\end{equation}

\noindent generates a Kodama-like time translation symmetry in Kerr-Vaidya spacetime.
\end{proposition}

\noindent \textbf{Proof.} For the vector field $K^a$ in \eqref{KodamaResult}, we directly observe that it is divergence-free, i.e. $\nabla_a K^a = 0$, and that it is tangent to surfaces of constant Boyer-Lindquist-$r$, i.e. 
\begin{equation}
K^a \nabla_a r = 0. 
\end{equation}

\noindent Moreover, we want to prove that the associated current $J^a = G^{ab}K_b$, constructed via the Einstein tensor $G^{ab}$, is covariantly conserved. Therefore, we consider the components of $J^a$, which are in ingoing coordinates $(v,r,\vartheta,\psi)$ explicitly given by \cite{Mathematica, xAct, xPerm, xTensor, xCoba}
\begin{align}
J^v &= \frac{1}{\rho^6} a^2 \sin^2\vartheta \left(a^2\cos^2\vartheta - r^2\right) M'(v) \label{Jv} ,\\
J^r &= \frac{1}{\rho^6} \left( \left( \rho^4 + r^4 + r^2 a^2 \sin^2\vartheta - a^4\cos^2\vartheta \right) M'(v) - r\rho^2 a^2 \sin^2\vartheta M''(v)\right) \label{Jr} ,\\
J^\vartheta &= \frac{1}{\rho^6} r a^2 \sin(2\vartheta) M'(v) \label{Jtheta},\\
J^\psi &= \frac{1}{\rho^6} a \left( a^2 \cos^2\vartheta - r^2 \right) M'(v). \label{Jpsi}
\end{align}

\noindent An explicit calculation \cite{Mathematica, xAct, xPerm, xTensor, xCoba} then shows that, indeed, 
\begin{equation}
\nabla_a J^a = 0.
\end{equation}

\noindent Thus, the vector field $K^a$ fulfills the same properties as the Kodama vector field $k^a$ in spherical symmetry. We conclude that $K^a$ generates a Kodama-like time translation symmetry in axisymmetric Kerr-Vaidya spacetime. $\qedsymbol$ \\

We remark that the Kodama-like vector field $K^a$ given in equation \eqref{KodamaResult} is not timelike in the entire Kerr-Vaidya spacetime. In fact, it becomes null when the $vv$-component of the metric \eqref{KerrVaidyaMetric} vanishes, namely at
\begin{equation}
K^a K_a = g_{vv} = -\left( 1 - \frac{2 M(v) r_E}{r_E^2 + a^2\cos^2\vartheta} \right) = 0 ,
\end{equation}

\noindent while $K^a$ becomes spacelike for $r_+ < r < r_E$. This is a direct analogy to the stationary Kerr metric, where the timelike Killing field $\left(\frac{\partial}{\partial t}\right)^a = \left(\frac{\partial}{\partial v}\right)^a$ \cite{MTW:1973grav} becomes spacelike (and also null) outside of the outer horizon $r_+$, hence, giving rise to the ergoregion, see \cite{Wald:1984gr}. In stationary Kerr spacetime, i.e. $M'(v)=0$, there exists the Killing vector field
\begin{equation}
\label{ZetaKillingField}
\zeta^a = \left( \frac{\partial}{\partial v} \right)^a + \frac{a}{r_+^2 + a^2} \left( \frac{\partial}{\partial \psi} \right)^a,
\end{equation}

\noindent which is also timelike \emph{inside} the ergoregion, and only becomes null on the outer horizon $r_+$ \cite{MTW:1973grav, Wald:1984gr}. A direct computation confirms that $\zeta^a$ does not generate a Kodama-like symmetry in Kerr-Vaidya spacetime, as the associated current $G^{ab}\zeta_b$ is not covariantly conserved for $M'(v) \neq 0$ \cite{Mathematica,xAct,xCoba,xPerm,xTensor}.

\begin{proposition}\label{KodamaUniqueness}
Let $\xi^a$ be a timelike Killing vector field in stationary Kerr spacetime, i.e.
\begin{equation}
\nabla_a \xi_b + \nabla_b \xi_a = 0,
\end{equation}

\noindent for $M'(v) = 0$, and $\xi^a \xi_a < 0$. Then, $\xi^a$ generalizes to a Kodama-like vector field in the whole exterior of the ergoregion in Kerr-Vaidya spacetime, if and only if $\xi^a$ is proportional to $K^a$.
\end{proposition}

\noindent \textbf{Proof.} An arbitrary timelike Killing vector field $\xi^a$ in stationary Kerr spacetime is given by the linear combination of Killing vector fields
\begin{equation}
\label{LinearCombination}
\xi^a = V \left( \frac{\partial}{\partial v} \right)^a + \Psi  \left( \frac{\partial}{\partial \psi} \right)^a,
\end{equation}

\noindent together with the timelike restriction \cite{Mathematica,xAct,xCoba,xPerm,xTensor}
\begin{equation}
\label{PsiInterval}
\frac{2 M(v) r a V - \frac{\rho^2}{\sin\vartheta} \sqrt{V^2 \Delta}}{\Gamma\rho^2 + 2M(v)r a^2\sin^2\vartheta} < \Psi < \frac{2 M(v) r a V + \frac{\rho^2}{\sin\vartheta} \sqrt{V^2 \Delta}}{\Gamma\rho^2 + 2M(v)r a^2\sin^2\vartheta},
\end{equation}

\noindent where $V,\Psi \in \mathbb{R}$ denote proportionality factors. Note that the lower bound in expression \eqref{PsiInterval} becomes positive within the ergoregion, i.e. $r < r_E$. Then, as a consequence of Proposition \ref{KodamaKV} and the property $\nabla_a \left( G^{ab} \phi_b \right) = 0$ for the Killing vector field $\phi^a$ \cite{Carroll:2004gr}, we conclude that the current $G^{ab} \xi_b$ is also covariantly conserved in the non-stationary Kerr-Vaidya geometry, as long as $V$ and $\Psi$ remain constant\footnote{We like to point out that this is not the case for the vector field $\zeta^a$ from expression \eqref{ZetaKillingField}, since $r_+$ depends on the advanced time $v$ in Kerr-Vaidya spacetime.}. Lastly, we use the requirement that $\xi^a$ needs to be timelike for all $r > r_E$.  The estimate \eqref{PsiInterval} shows that $\Psi$ must vanish in the limit $r\rightarrow\infty$. $\qedsymbol$ \\

The statement of Proposition \ref{KodamaUniqueness} seems rather obvious since $\left( \frac{\partial}{\partial v} \right)^a$ is the only Killing vector field in stationary Kerr spacetime which is timelike in the entire exterior of the ergosphere, as we have essentially proven above. Nonetheless, it further substantiates the interpretation of the Kodama-like vector field $K^a$ as the generalization of a timelike Killing vector field, and therefore as the preferred time evolution vector field, for Kerr-Vaidya spacetime.

\subsection{Noether Charges}
Viewing the current $J^a$ as a Noether current, we now investigate the corresponding charge $Q_K$ and its physical properties. Explicitly, the charge $Q_K$  of the conserved current $J^a$ is given by the integral over hypersurfaces $\Sigma_r$ of constant Boyer-Lindquist-$r$, which are defined by the normal co-vector $n_a = (dr)_a$ \cite{Hayward:1996sphsym}, i.e.
\begin{equation}
\label{ChargeIntegral}
Q_K = \frac{1}{8\pi} \int_{\Sigma_r} J^a n_a \, d\textrm{vol}_{\Sigma_r}.
\end{equation}

\noindent We would like to remark that the factor of $(8\pi)^{-1}$ in equation \eqref{ChargeIntegral} arises from the definition of the current $J^a$ via the Einstein tensor $G_{ab}$. In the literature, sometimes the stress energy tensor $T_{ab}$ is used instead, in which case the factor of $(8\pi)^{-1}$ is already included in the current.

\begin{proposition}\label{PropMassCharge}
In Kerr-Vaidya spacetime, the charge $Q_K$, corresponding to the Noether current $J^a$ generated by the Kodama-like vector field $K^a$, is given by
\begin{equation}
\label{KodamaCharge}
Q_K = M(v) - M'(v) \left( \frac{r^2 + a^2}{2a} \arctan\left(\frac{a}{r}\right) - \frac{r}{2} \right),
\end{equation}

\noindent which coincides with the Brown-York mass in the asymptotically flat region, i.e.
\begin{equation}
\label{AsymptoticCharge}
\lim_{r\rightarrow\infty} Q_K = M(v).
\end{equation}
\end{proposition}

\noindent The proof of this statement is given in Appendix \ref{ProofMassCharge}. \\

Note that in close proximity to the black hole, there is an additional contribution to the charge $Q_K$ by the advanced time derivative $M'(v)$ of the mass. If we interpret $Q_K$ as some form of quasi-local energy, this would mean that for evaporating black holes, i.e. $M'(v)<0$, this energy would be observed to be slightly larger than the energy of a stationary black hole, by an observer sufficiently close to the black hole. \\

Furthermore, viewing our calculation of the charge $Q_K$ as analogous to the computation of the Komar mass, see \cite{Komar:1959con, Vollick:2023mgr}, we conclude that in a similar fashion, we are able to calculate the angular momentum of Kerr-Vaidya spacetime as the charge $Q_\phi$ of a Noether current $I^a$ generated by the angular Killing vector field $\phi^a$.

\begin{corollary}\label{CorollaryAngularMomentum}
Let $\phi^a$ be the angular Killing vector field, as before, and $I^a = G^{ab}\phi_b$ the corresponding covariantly conserved current. Then, the charge $Q_\phi$ of $I^a$ is given by
\begin{equation}
\label{AngularMomentumFunction}
Q_\phi = M(v) a  - M'(v)\left( \frac{(r^2 + a^2)^2}{2a^2}\arctan\left( \frac{a}{r} \right) - \frac{r^3}{2a} - \frac{5ar}{6} \right),
\end{equation}

\noindent which in the asymptotically flat region converges to the angular momentum $L_\textrm{Kerr}$ of the stationary Kerr metric, i.e.
\begin{equation}
\label{AsymptoticAngularMomentum}
\lim_{r\rightarrow\infty} Q_\phi = M(v) a \equiv L_\textrm{Kerr}.
\end{equation}
\end{corollary}

\noindent The proof of this statement is given in Appendix \ref{ProofAngularMomentum}. \\

Analogously to the charge $Q_K$ of the Kodama-like current $J^a$, the charge $Q_\phi$ converges in the asymptotically flat region to the same expression for the angular momentum as in the stationary case. For an observer at significant proximity to the black hole, the observed angular momentum of the black hole contains an additional contribution which depends on $M'$. Again, for evaporating black holes, i.e. $M'(v)<0$, this additional term is positive. \\

We want to remark that also in spherical symmetry, there are instances where the charge of the Kodama current does not coincide with the Brown-York mass. We illustrate an example of this phenomenon in Appendix \ref{VaidyaBonnor}. \\

Altogether, we observe that the charge $Q_K$ or the Komar mass, respectively, coincide with the mass $M(v)$ in the asymptotically flat region of the spacetime. Similarly, the charge $Q_\phi$ or the Komar angular momentum coincides with the expression $L = M(v)a$ from the stationary setting in the asymptotically flat region. This may be closely related to the argument that, in dynamical spacetimes, quantities like mass or angular momentum should be evaluated in the asymptotically flat region\footnote{provided that it exists}, where the generators of corresponding symmetry become Killing vector fields, see \cite[Chapter 11]{Wald:1984gr}. Indeed, we can show by direct calculation that the Kodama-like vector field $K^a$ fulfills the Killing equation in the asymptotically flat region, i.e.
\begin{equation}
\lim_{r\rightarrow\infty} \left(\nabla_a K_b + \nabla_b K_a\right) = 0.
\end{equation}

In conclusion, we have found a Kodama-like vector field $K^a$ which fulfills the same properties as the Kodama vector field $k^a$ in spherical symmetry, including the generation of a covariantly conserved current $J^a$. The corresponding charge $Q_K$ coincides with the Brown-York mass $M(v)$ in the asymptotically flat region. Furthermore, we observe that $K^a$ shows the same causal behaviour as the timelike Killing vector field in stationary Kerr spacetime. Therefore, we argue that we have indeed constructed a Kodama-like vector field outside of spherical symmetry, namely, for Kerr-Vaidya spacetime.

\section{Generalization to a Cosmological Background}\label{CosmologicalGeneralization}
In the next step, we aim to generalize Kerr-Vaidya spacetime to a de Sitter background with a non-vanishing cosmological constant $\Lambda \neq 0$. This leads us to the qualitatively different Kerr-Vaidya-de Sitter metric, which we propose and investigate in this section. (To our knowledge, this is an original proposal for a spacetime of Kerr-Vaidya-de Sitter type.)

\subsection{Kerr-Vaidya-de Sitter Spacetime}\label{KVdS}
Given the stationary Kerr metric, there exists a corresponding generalization to a de Sitter background, namely, the Kerr-de Sitter metric, see \cite{AM:2011kds, LZ:2015kds}. This metric is a solution to the Einstein equations with a non-zero cosmological constant $\Lambda \neq 0$, and is for Boyer-Lindquist-type coordinates $(t,r,\vartheta,\varphi)$ explicitly given by the line element \cite{Borthwick:2018kds}
\begin{align}
ds^2 = &-\frac{\Delta_\Lambda - \Theta a^2\sin^2\vartheta}{\rho^2 Z^2} dt^2 - \frac{2a \sin^2 \vartheta}{\rho^2 Z^2}\left( \Gamma\Theta - \Delta_\Lambda \right) dt d\varphi + \frac{\rho^2}{\Delta_\Lambda} dr^2 + \frac{\rho^2}{\Theta} d\vartheta^2 \nonumber \\
&+ \frac{\sin^2 \vartheta}{\rho^2 Z^2} \left( \Gamma^2 \Theta - \Delta_\Lambda a^2 \sin^2\vartheta \right) d\varphi^2.
\label{standardKdS}
\end{align}

\noindent Here, additionally to the previous definitions, we have used the shorthand notation
\begin{align}
\Delta_\Lambda &= \Delta - \Gamma r^2 H^2 , \\
\Theta &= 1 + H^2 a^2 \cos^2 \vartheta , \\
Z &= 1 + H^2 a^2,
\end{align}

\noindent where $H$ denotes the Hubble constant, which is related to the cosmological constant $\Lambda$ via the relation \cite{Weinberg:1972gr}
\begin{equation}
H = \sqrt{\frac{\Lambda}{3}}.
\end{equation}

In order to find the non-stationary analogue to the Kerr-de Sitter metric, we express the metric \eqref{standardKdS} in terms of ingoing Eddingtion-Finkelstein-type coordinates $(\tilde{v},r,\vartheta,\tilde{\psi})$, defined by the transformations \cite{Borthwick:2018kds}
\begin{align}
d\tilde{v} &= dt + \frac{\Gamma Z}{\Delta_\Lambda} dr , \\
d\tilde{\psi} &= d\varphi + \frac{a Z}{\Delta_\Lambda} dr.
\end{align}

\noindent In analogy to the previous section, we then assign dynamics to the mass $M$ by replacing it with a function $M(\tilde{v})$ of the new advanced time $\tilde{v}$. Altogether, the resulting Kerr-Vaidya-de Sitter metric takes the explicit form
\begin{align}
ds^2 = &-\frac{\Delta_\Lambda - \Theta a^2\sin^2\vartheta}{\rho^2 Z^2} d\tilde{v}^2 + \frac{2}{Z}d\tilde{v}dr - \frac{2a \sin^2 \vartheta}{\rho^2 Z^2}\left( \Gamma\Theta - \Delta_\Lambda \right) d\tilde{v} d\tilde{\psi} - \frac{2a\sin^2\vartheta}{Z} dr d\tilde{\psi} \nonumber \\
&+ \frac{\rho^2}{\Theta} d\vartheta^2 + \frac{\sin^2 \vartheta}{\rho^2 Z^2} \left( \Gamma^2 \Theta - \Delta_\Lambda a^2 \sin^2\vartheta \right) d\tilde{\psi}^2.
\label{KerrVaidyadSMetric}
\end{align}

Examining the properties of the Kerr-Vaidya-de Sitter metric, we directly observe that the spacelike vector field $\phi^a$, given in expression \eqref{AxialKillingField}, is again a Killing field. Hence, Kerr-Vaidya-de Sitter spacetime remains to be axisymmetric. Furthermore, we find that similar to the previous case, we obtain the square root of the (negative) determinant of the metric
\begin{equation}
\sqrt{-\tilde{g}} = \frac{\rho^2 \sin\vartheta}{Z^2}.
\end{equation}

\noindent However, in contrast to the Kerr-Vaidya metric, the Kerr-Vaidya-de Sitter metric has a non-zero Ricci scalar, given in terms of the cosmological constant $\Lambda$ by
\begin{equation}
\tilde{R} = 4\Lambda.
\end{equation}

\noindent Moreover, with respect to the ingoing coordinate system, the Einstein tensor $\tilde{G}_{ab}$ takes the schematic form
\begin{equation}
\label{KVdSET}
(\tilde{G}_{\mu\nu}) = \begin{pmatrix}
\tilde{G}_{\tilde{v}\tilde{v}} & -\frac{\Lambda}{Z} & \tilde{G}_{\tilde{v}\vartheta} & \tilde{G}_{\tilde{v}\tilde{\psi}} \\
-\frac{\Lambda}{Z} & 0 & 0 & \frac{\Lambda}{Z} a \sin^2 \vartheta \\
\tilde{G}_{\vartheta \tilde{v}} & 0 & \tilde{G}_{\vartheta\vartheta} & \tilde{G}_{\vartheta\tilde{\psi}} \\
\tilde{G}_{\tilde{\psi} \tilde{v}} & \frac{\Lambda}{Z} a \sin^2 \vartheta & \tilde{G}_{\tilde{\psi} \vartheta} & \tilde{G}_{\tilde{\psi}\tilde{\psi}}
\end{pmatrix}  ,
\end{equation}

\noindent with the $\tilde{v}\tilde{v}$-component explicitly given by \cite{Mathematica, xAct, xPerm, xTensor, xCoba}
\begin{align}
\tilde{G}_{\tilde{v}\tilde{v}} &= \frac{\Lambda}{Z^2} \left( \Theta - \frac{\Gamma\Lambda}{3} - \frac{2 M(\tilde{v}) r}{\rho^2} \right) \\
&\hspace{0,5cm} + \frac{2}{\rho^6}\left( \frac{r^4 + r^2 a^2 - a^4 \sin^2\vartheta \cos^2 \vartheta}{Z} M'(\tilde{v}) - \frac{\rho^2 r a^2 \sin^2 \vartheta}{2\Theta} M''(\tilde{v}) \right). \nonumber
\end{align}

\noindent Note that in the asymptotically flat limit, i.e. $\Lambda\rightarrow 0$, we retrieve the respective results for the Kerr-Vaidya metric from Section \ref{KV}. Thus, the Kerr-Vaidya-de Sitter metric also violates all of the standard energy conditions, so that it is mostly relevant for quantum effects. \\

Considering the horizons of stationary Kerr-de Sitter spacetime, the primary difference to Kerr spacetime is the existence of a cosmological horizon $r_C$, which is due to the equation $\Delta_\Lambda = 0$ having four distinct real solutions for $\Lambda \neq 0$ and $a<M$ \cite{Borthwick:2018kds}. Three of these four solutions correspond to the inner, the outer, and the cosmological horizon, respectively, i.e.
\begin{equation}
\Delta_\Lambda (r_-) = \Delta_\Lambda (r_+) = \Delta_\Lambda (r_C) = 0,
\end{equation}

\noindent with the fourth solution usually to be considered unphysical since it is either negative or complex for sub-extremal Kerr-de Sitter spacetimes, see \cite{Borthwick:2018kds}. \\

In particular, this means that the black hole horizons and the cosmological horizons are \emph{not} independent from each other, since both depend on the mass $M$ and the cosmological constant $\Lambda$ \cite{LMZZ:2017kds}. Hence, we conclude that in the dynamical case of Kerr-Vaidya-de Sitter spacetime, the dynamical mass of the black hole likely induces a dynamical behaviour of the cosmological horizon as well. Since the localization of the different horizons is expected to be even more challenging than in the asymptotically flat case, we leave this problem for future investigations.

\subsection{Kodama-like Symmetry in Kerr-Vaidya-de Sitter Spacetime}\label{KodamaKVdS}
Next, we want to extend our results from Section \ref{KodamaLike} to Kerr-Vaidya-de Sitter spacetime. In this generalized setting, additional properties of the spacetime have to be taken into account, most importantly, the asymptotic de Sitter behaviour of the metric, instead of asymptotic flatness, see \cite{ABK:2015dS}. This poses various additional challenges for the construction of the Kodama-like symmetry and its corresponding quantities. Nonetheless, we are still able to perform an analogous computation \cite{Mathematica, xAct, xCoba, xPerm, xTensor} to obtain that the analogue to the Kodama-like vector field in Proposition \ref{KodamaKV} also exhibits Kodama-like behaviour in the generalized Kerr-Vaidya-de Sitter geometry.

\begin{proposition}
The vector field
\begin{equation}
\tilde{K}^a = \left( \frac{\partial}{\partial \tilde{v}} \right)^a
\end{equation}

\noindent generates a Kodama-like time translation symmetry in Kerr-Vaidya-de Sitter spacetime.
\end{proposition}

In particular, the Kodama-like vector field $\tilde{K}^a$ is divergence-free, tangent to surfaces of constant $r$, and gives rise to a covariantly conserved current $\tilde{J}^a = \tilde{G}^{ab}\tilde{K}_b$. Furthermore, it possesses a causal behaviour similar to the Kodama-like vector field in the asymptotically flat case, giving rise to an ergo-like region, in the sense that $\tilde{K}^a\tilde{K}_a = 0$ does not coincide with $\Delta_\Lambda = 0$ \cite{Mathematica, xAct, xCoba, xPerm, xTensor}. \\

The associated charge $Q_{\tilde{K}}$ to the Noether current $\tilde{J}^a$ is again obtained via the integral from expression \eqref{ChargeIntegral}. As before, we obtain two terms, one of which computes to be the dynamical mass $M(\tilde{v})$ (times a constant factor), while the second term of the integral is again proportional to $M'(\tilde{v})$. The same argument holds for the computation of the angular momentum, analogous to Corollary \ref{CorollaryAngularMomentum}. The computation and a brief discussion of these results is illustrated in Appendix \ref{IntegralsKVdS}.

\section{Conclusions \& Outlook}
In this work, we have presented a Kodama-like time translation symmetry for Kerr-Vaidya spacetime. This symmetry is generated by a (mostly) timelike and divergence-free vector field which is tangent to surfaces of constant Boyer-Lindquist-$r$ and, most importantly, gives rise to a covariantly conserved current when contracted with the Einstein tensor. Furthermore, we have computed the Noether charge corresponding to the Kodama-like current. We find that the charge converges to the spacetime's Brown-York mass $M(v)$ in the asymptotically flat region. For an observer sufficiently close to the Kerr-Vaidya black hole, the charge contains additional terms, which depend linearly on the (negative) advanced time derivative $M'(v)$ of the mass. \\

Additionally, we have proved explicitly that in Kerr-Vaidya-spacetime, the Brown-York quasi-local mass is given by $M(v)$. This means that the usual physical interpretation of $M(v)$ as the dynamical mass of the spacetime, see \cite{DT:2020kv,Dahal:2021kvh,DMST:2023KVii}, is indeed justified. Moreover, we have computed the angular momentum of Kerr-Vaidya spacetime as the charge of the Noether current induced by the angular Killing vector field $\phi^a$. It turns out that the angular momentum converges to the same expression as the angular momentum of stationary Kerr spacetime in the asymptotically flat region. \\

In Section \ref{CosmologicalGeneralization}, we proposed a generalization of the Kodama-like symmetry to a Kerr-Vaidya-de Sitter-type spacetime by showing that a vector field with the necessary properties still exists in the cosmological background. We found that the Kodama-like vector field and the associated charges in Kerr-Vaidya-de Sitter spacetime show the same qualitative behaviour as their counterpart in the asymptotically flat case. The evaluation of the charges for distant observers, however, proves to be substantially more challenging, due to the asymptotic behaviour of the spacetime, and will be left for future work. \\

One limitation for the Kodama-like symmetry from Sections \ref{KodamaLike} and \ref{KodamaKVdS} is that it does not necessarily apply to general axisymmetric dynamical spacetimes. For example, if we allow for a dynamical rotation parameter $a(v)$, the currents corresponding to the vector fields $K^a$ and $\tilde{K}^a$, respectively, are no longer covariantly conserved. Thus, we conclude that the Kodama-like symmetry presented here is only valid for $a'(v) = 0$. It would be of interest for future investigations to work out the most general class of axisymmetric spacetimes which possess such a Kodama-like time translation symmetry. \\

Furthermore, we would like to point out that the usual spherically symmetric relation for the surface gravity associated to the Kodama vector field, see \cite{Faraoni:2015horizons}, does not hold for the Kodama-like symmetry in axisymmetry. Here, further considerations are required, in oder to find a suitable replacement for the surface gravity associated to the Kodama-like vector fields. We refer to \cite{MH:2000dynsg} and \cite{NY:2008dynsg} for discussions of generalized notions of surface gravities. \\

Finally, we briefly comment on the possibility of application of our results to future works. As mentioned before, the Kodama symmetry in spherically symmetric spacetimes has been a very useful tool in the field of semi-classical gravity, e.g. to study the back reaction of quantum effects on stationary black hole spacetimes, or to investigate quantum field theory in spherically symmetric dynamical backgrounds, see \cite{KPV:2021ea,DAngelo:2021bhre}. It would be fascinating to study, to which extent our results enable a first generalization of well-established results in spherical symmetry to axisymmetric spacetimes.

\section*{Acknowledgements}
We express our gratitude to Valerio Faraoni for very productive discussions about Kodama vector fields, and Christiane Klein for her expertise and ideas on rotating black holes in de Sitter backgrounds. Furthermore, we would like to thank Maria Alberti, Marc Casals, Stefan Czimek, Markus Fr\"ob, Daniel Fr\"ohlich, Dejan Gajic, Daan Janssen, José Senovilla, and Daniel Terno for inspiring conversations about various details. We also thank Thiago Campos, Jochen Zahn, and and two anonymous referees for very helpful comments on a previous version of this work. \\

The work of the first named author (PD) has been funded by the Deutsche Forschungsgemeinschaft (DFG) under Grant No.\ 406116891 within the Research Training Group (RTG) 2522: “Dynamics and Criticality in Quantum and Gravitational Systems”. The work of the second named author (RV) has received funding by the CY Initiative of Excellence (grant "Investissements d'Avenir" ANR-16-IDEX-0008) during his stay at the CY Advanced Studies, whose support is gratefully acknowledged.

\appendix
\section{Proofs}
In the first appendix, we provide the detailed proofs that are too extensive for the main text of our work.

\subsection{Proof of Theorem \ref{Theorem_BY}}\label{Proof_BY}
Firstly, we consider the definition of the Brown-York quasi-local mass, given by
\begin{equation}
\label{BrownYorkMass}
M_\textrm{BY} = \frac{1}{8\pi}\int_{S_r} \left( \mathcal{K}_0 - \mathcal{K}_g \right) d\textrm{vol}_{S_r},
\end{equation}

\noindent where $S_r$ is a closed 2-dimensional surface of constant Boyer-Lindquist-$r$, $\mathcal{K}_g$ is the trace of the extrinsic curvature of $S_r$ embedded in Kerr-Vaidya spacetime, and $\mathcal{K}_0$ denotes the trace of the extrinsic curvature of $S_r$ embedded isometrically in flat spacetime \cite{BY:1993mass, Poisson:2004tool, MTXM:2010by}. Given a normal co-vector $n_a = (dr)_a$ to $S_r$, the trace $\mathcal{K}_g$ is computed by \cite{Poisson:2004tool, Mathematica, xAct, xPerm, xTensor, xCoba}
\begin{equation}
\mathcal{K}_g = \nabla_a n^a = -\frac{2\left( M(v) - r \right)}{\rho^2}.
\end{equation}

\noindent In order to compute the extrinsic curvature of the ellipsoidal surfaces $S_r$ embedded in flat spacetime, we consider the Minkowski metric in Cartesian coordinates $(t,x,y,z)$ and apply the coordinate transformation
\begin{align}
x &= \rho \sin\vartheta \cos\varphi ,\\
y &= \rho \sin\vartheta \sin\varphi ,\\
z &= r\cos\vartheta ,
\end{align}

\noindent to \emph{oblate spheroidal} coordinates $(t,r,\vartheta,\varphi)$, resulting in the line element \cite{PI:1998flatkerr, Visser:2009kerr}
\begin{equation}
ds^2 = -dt^2 + \frac{\rho^2}{\Gamma} dr^2 + \rho^2 d\vartheta^2 + \Gamma \sin^2\vartheta d\varphi^2.
\end{equation}

\noindent Note that we have used the same abbreviations $\rho^2$ and $\Gamma$ as throughout the paper. A transformation $d\overline{v}=dt+dr$ and $d\psi = d\varphi + \frac{a}{\Gamma}dr$ to ingoing Eddington-Finkelstein-type coordinates then leads us to a metric for Minkowski spacetime, which corresponds to the massless limit of the Kerr-Vaidya metric. In particular, it holds for the volume element of both metrics that
\begin{equation}
\sqrt{-g} = \rho^2 \sin\vartheta.
\end{equation}

\noindent Thus, we are able to isometrically embed the surfaces of constant Boyer-Lindquist-$r$ in Minkowski spacetime, using the coordinate system $(\overline{v},r,\vartheta,\psi)$. Given the normal co-vector $n_a = (dr)_a$ in flat spacetime, we analogously compute the trace of the extrinsic curvature
\begin{equation}
\mathcal{K}_0 = \overline{\nabla}_a n^a = \frac{2r}{\rho^2},
\end{equation}

\noindent where $\overline{\nabla}$ denotes the covariant derivative with respect to the flat metric. Altogether, we therefore obtain the Brown-York quasi-local mass
\begin{align}
M_\textrm{BY} &= \frac{1}{8\pi} \int_{S_r} \left( \frac{2r}{\rho^2} + \frac{2\left( M(v) - r \right)}{\rho^2} \right) d\textrm{vol}_{S_r} \nonumber \\
&= \frac{1}{4\pi} \int_0^{2\pi} d\psi \int_0^\pi d\vartheta \rho^2 \sin\vartheta \frac{M(v)}{\rho^2} \nonumber \\
&= \frac{M(v)}{2} \int_0^\pi \sin\vartheta d\vartheta \nonumber \\
&= M(v),
\end{align}

\noindent which is the required result.

\subsection{Proof of Proposition \ref{PropMassCharge}}\label{ProofMassCharge}
Let us recall the components of the Kodama-like current $J^a$, given in expressions \eqref{Jv}-\eqref{Jpsi}. Given the normal co-vector $n_a = (dr)_a$, the integrand from equation \eqref{ChargeIntegral} reduces to
\begin{equation}
J^a n_a = \frac{1}{\rho^6} \left( \left( \rho^4 + r^4 + r^2 a^2 \sin^2\vartheta - a^4\cos^2\vartheta \right) M'(v) - r\rho^2 a^2 \sin^2\vartheta M''(v)\right).
\end{equation}

\noindent Together with the volume form $d\textrm{vol}_{\Sigma_r} = \rho^2\sin\vartheta\, dv \, d\vartheta \, d\psi$, we therefore obtain the Noether charge
\begin{equation}
\label{ExplicitChargeIntegral}
Q_K = \frac{1}{8\pi} \int_{\Sigma_r} \frac{\sin\vartheta}{\rho^4} \left( \left( \rho^4 + r^4 + r^2 a^2 \sin^2\vartheta - a^4\cos^2\vartheta \right) M'(v) - r\rho^2 a^2 \sin^2\vartheta M''(v)\right) dv \, d\vartheta \, d\psi.
\end{equation}

\noindent Next, we use that $M(v)$ at a given null coordinate $v$ is given by the integral
\begin{equation}
M(v) = \int_{-\infty}^v M'(\underline{v}) d\underline{v},
\end{equation}

\noindent see also \cite[Appendix B]{Hayward:1996sphsym}, where we have set the boundary condition $\lim_{v\rightarrow -\infty}M(v) = 0$, corresponding to a physical black hole that forms from gravitational collapse. Hence, we may reformulate expression \eqref{ExplicitChargeIntegral} as
\begin{align}
Q_K &= \frac{1}{8\pi} \left[ 2\pi M(v) \int_0^\pi \sin\vartheta d\vartheta + 2\pi (r^2 + a^2) M(v) \int_0^\pi \frac{r^2 - a^2 \cos^2\vartheta}{(r^2 + a^2\cos^2\vartheta)^2} \sin\vartheta d\vartheta \right. \nonumber \\
&\hspace{1.1cm} - \left. 2\pi a^2 r M'(v) \int_0^\pi \frac{\sin^3\vartheta}{r^2 + a^2\cos^2\vartheta}d\vartheta \right],
\label{ThreeThetaIntegrals}
\end{align}

\noindent where we have already carried out the integral over $\psi$. While the first integral with respect to $\vartheta$ is trivial, the second integral can be solved by the substitution $y = \cos\vartheta$, so that we obtain 
\begin{equation}
\label{SecondIntegral}
\int_0^\pi \frac{r^2 - a^2 \cos^2\vartheta}{(r^2 + a^2\cos^2\vartheta)^2} \sin\vartheta d\vartheta = - \int_{-1}^1 \frac{a^2 y^2 - r^2}{(a^2 y^2 + r^2)^2} dy.
\end{equation}

\noindent Using some manipulations, we can further split this integral into two parts of the form
\begin{equation}
- \int_{-1}^1 \frac{a^2 y^2 - r^2}{(a^2 y^2 + r^2)^2} dy = - \int_{-1}^1 \frac{dy}{a^2 y^2 + r^2} + 2r^2 \int_{-1}^1 \frac{dy}{(a^2 y^2 + r^2)^2}.
\end{equation}

\noindent Here, the integral in the first term can be solved by another substitution $w = \frac{a y}{r}$ and computes as
\begin{equation}
\label{ArcTanIntegralOne}
\int \frac{dy}{a^2 y^2 + r^2} = \frac{\arctan\left( \frac{ay}{r} \right)}{ar}.
\end{equation}

\noindent In order to solve the integral in the second term, we need to make use of the reduction formula for $n\in\mathbb{N}$ \cite{Gradshteyn:1980int}
\begin{equation}
\int \frac{dx}{(\alpha x^2 + \beta)^n} = \frac{x}{2\beta (n-1) (\alpha x^2 + \beta)^{n-1}} + \frac{2n-3}{2\beta (n-1)} \int \frac{dx}{(\alpha x^2 + \beta)^{n-1}},
\end{equation}

\noindent so that we obtain 
\begin{align}
\int \frac{dy}{(a^2 y^2 + r^2)^2} &= \frac{y}{2 r^2 (a^2 y^2 + r^2)} + \frac{1}{2r^2} \int \frac{dy}{a^2 y^2 + r^2} \\
&= \frac{y}{2 r^2 (a^2 y^2 + r^2)} + \frac{\arctan\left( \frac{ay}{r}\right)}{2ar^3}.
\end{align}

\noindent Altogether, the terms containing the $\arctan$ function cancel out and the integral in \eqref{SecondIntegral} finally reduces to
\begin{equation}
- \int_{-1}^1 \frac{a^2 y^2 - r^2}{(a^2 y^2 + r^2)^2} dy = \left. \frac{y}{a^2 y^2 + r^2}\right\vert_{-1}^1 = \frac{2}{r^2 + a^2}.
\end{equation}

\noindent Proceeding to the third integral in equation \eqref{ThreeThetaIntegrals}, we observe that we are able to use the same substitutions as before, in order to write
\begin{equation}
\int_0^\pi \frac{\sin^3\vartheta}{r^2 + a^2\cos^2\vartheta}d\vartheta = \int_{-1}^1 \frac{y^2 -1}{a^2 y^2 + r^2} dy.
\end{equation}

\noindent Utilizing the previous computations, we ultimately arrive at
\begin{equation}
\int_{-1}^1 \frac{y^2 -1}{a^2 y^2 + r^2} dy = \left. \frac{ (r^2 + a^2) \arctan\left( \frac{ay}{r}\right)}{a^3r} - \frac{y}{a^2} \right\vert_{-1}^1 = \frac{ 2(r^2 + a^2) \arctan\left( \frac{a}{r}\right)}{a^3r} - \frac{2}{a^2}.
\end{equation}

\noindent Inserting our results into \eqref{ThreeThetaIntegrals} directly leads to the charge of the Kodama-like vector field stated in equation \eqref{KodamaCharge}. \\

Lastly, we need to prove that the function
\begin{equation}
q(r) = \frac{r^2+a^2}{2a}\arctan\left(\frac{a}{r}\right) - \frac{r}{2}
\end{equation}

\noindent converges to zero for for $r\rightarrow\infty$. Therefore, we use that for sufficiently large $r$, i.e. sufficiently small arguments $x$, we can use the linear expansion $\arctan(x)\approx x$. Hence, for sufficiently large $r$, the function $q(r)$ altogether behaves as
\begin{equation}
q(r) \longrightarrow \frac{r^2 + a^2}{2a} \cdot \frac{a}{r} - \frac{r}{2} = \frac{a^2}{2r},
\end{equation}

\noindent which converges to zero in the asymptotically flat region, i.e.
\begin{equation}
\lim_{r\rightarrow\infty} \, \frac{a^2}{2r} = 0.
\end{equation}

\noindent Note that higher order terms of the polynomial expansion, which we have used above, converge to zero for $r\rightarrow \infty$, as well.

\subsection{Proof of Corollary \ref{CorollaryAngularMomentum}}\label{ProofAngularMomentum}
In order to obtain the charge of the current generated by the spacelike Killing vector field $\phi^a$, we begin with the integral
\begin{equation}
    Q_\phi = - \frac{1}{8\pi} \int_{\Sigma_r} I^a n_a d\textrm{vol}_{\Sigma_r}
\end{equation}

\noindent over surfaces $\Sigma_r$ of constant $r$ with normal $n_a = (dr)_a$. Analogously to the calculation in Appendix \ref{ProofMassCharge}, we obtain the expression
\begin{align}
Q_\phi &= -\frac{1}{8\pi} \left[ \int_{\Sigma_r} \frac{a \sin^3\vartheta \left( a^2 (a^2\cos^2\vartheta - r^2) - r^2 (\rho^2 + 2r^2) \right)}{\rho^4} M'(v) \, dv \, d\vartheta \, d\psi \right. \nonumber \\
&\hspace{1,73cm} + \left. \int_{\Sigma_r} \frac{a^3 r \sin^5\vartheta}{\rho^2} M''(v) \, dv\, d\vartheta \, d\psi \right],
\end{align}

\noindent which we can reformulate to 
\begin{align}
Q_\phi &= -\frac{1}{4} \left[ M(v) \left( \int_0^\pi \frac{a^3 \sin^3\vartheta (a^2\cos^2\vartheta - r^2)}{(r^2 + a^2\cos^2\vartheta)^2}d\vartheta - \int_0^\pi \frac{ar^2 \sin^3\vartheta}{r^2 + a^2\cos^2\vartheta} d\vartheta \right. \right. \nonumber \\
&\hspace{1,62cm} \left. \left. -\int_0^\pi \frac{2ar^4 \sin^3\vartheta}{(r^2 + a^2\cos^2\vartheta)^2} d\vartheta \right) + M'(v) \int_0^\pi \frac{a^3 r \sin^5\vartheta}{r^2 + a^2\cos^2\vartheta} d\vartheta \right],
\label{FourThetaIntegrals}
\end{align}

\noindent where we have carried out the integrals over $v$ and $\psi$, as before. Note that we have already solved the second integral above in the previous proof, so that we can write
\begin{equation}
a r^2 \int_0^\pi \frac{\sin^3 \vartheta}{r^2 + a^2\cos^2\vartheta} d\vartheta = \frac{2r (r^2 + a^2)}{a^2}\arctan\left(\frac{a}{r}\right) - \frac{2r^2}{a}.
\end{equation}

\noindent In a very similar computation, we are able to solve the third integral in equation \eqref{FourThetaIntegrals}, in order to to obtain
\begin{equation}
2a r^4 \int_0^\pi \frac{\sin^3 \vartheta}{(r^2 + a^2\cos^2\vartheta)^2} d\vartheta = \frac{2r (a^2 - r^2)}{a^2}\arctan\left(\frac{a}{r}\right) + \frac{2r^2}{a}.
\end{equation}

\noindent Next, let us consider the first integral from expression \eqref{FourThetaIntegrals}. By the substitutions $y = \cos\vartheta$ and $w = \frac{ay}{r}$, we are able to solve this integral in the same manner as in the proof of Proposition \ref{PropMassCharge}, so that we ultimately obtain
\begin{equation}
a^3 \int_0^\pi \frac{\sin^3\vartheta (a^2 \cos^2\vartheta - r^2)}{(r^2 + a^2\cos^2\vartheta)^2} d\vartheta = 4r\arctan\left(\frac{a}{r}\right) - 4a.
\end{equation}

\noindent Inserting the results for the first three integrals, we notice that most terms cancel out, so that expression \eqref{FourThetaIntegrals} simplifies to
\begin{equation}
Q_\phi = -\frac{1}{4} \left( -4a M(v) + M'(v)\int_0^\pi \frac{a^3 r \sin^5\vartheta}{r^2 + a^2\cos^2\vartheta}d\vartheta \right).
\end{equation}

\noindent Lastly, we need to solve the remaining integral, for which we can use the same methods and substitutions as in the previous calculations. We obtain
\begin{equation}
a^3 r \int_0^\pi \frac{\sin^5\vartheta}{r^2 + a^2\cos^2\vartheta}d\vartheta = \frac{2}{a^2}\left((r^2 + a^2)^2 \arctan\left(\frac{a}{r}\right) - ar^3 \right) - \frac{10}{3}ar.
\end{equation}

\noindent Combining our results, we directly obtain the charge $Q_\phi$ given by expression \eqref{AngularMomentumFunction}. \\

Moreover, in order to prove equation \eqref{AsymptoticAngularMomentum} for the charge in the asymptotically flat region, we need to examine the function
\begin{equation}
\ell (r) = \frac{(r^2 + a^2)^2}{2a^2}\arctan\left( \frac{a}{r} \right) - \frac{r^3}{2a} - \frac{5ar}{6}
\end{equation}

\noindent and its behaviour for $r\rightarrow\infty$. Therefore, we consider the extension of $\ell(r)$ to the complex plane, using that the $\arctan$ function is for $z\in\mathbb{C}$ given by
\begin{equation}
\arctan(z) = \frac{\ln(1+iz) - \ln(1-iz)}{2i}.
\end{equation}

\noindent Then, we consider the corresponding Laurent expansion about $z=\infty$, given by the alternating series \cite{Mathematica}
\begin{equation}
\ell(z) = \frac{4a^3}{15z} - \frac{4a^5}{105z^3} + \frac{4a^7}{315z^5} - \frac{4a^9}{693z^7} + \frac{4a^{11}}{1287z^9} - \cdots,
\end{equation}

\noindent which we can rewrite as
\begin{equation}
\ell (z) = 4a^2 \sum_{n=0}^\infty \frac{(-1)^n}{(2n+1)(2n+3)(2n+5)}\left(\frac{a}{z}\right)^{2n+1}.
\end{equation}

\noindent Restricting this series to the real numbers again, i.e. $\textrm{Im}z = 0$, and using the dominated convergence theorem for large $r = \textrm{Re}z$, we observe that the series vanishes in the limit of $r\rightarrow\infty$, which concludes our proof. We remark that this route of proving convergence could have also been used to prove Proposition \ref{PropMassCharge}, and vice versa.

\section{Vaidya-Bonnor Spacetime}\label{VaidyaBonnor}
In this appendix we shortly discuss the Vaidya-Bonnor metric, a non-stationary solution to the Einstein-Maxwell equations, which serves as a useful spherically symmetric toy model for Kerr-Vaidya-type spacetimes. It is, in ingoing Eddington-Finkelstein coordinates $(v,r,\vartheta,\varphi)$, given by the line element
\begin{equation}
ds^2 = -\left( 1 - \frac{2M(v)}{r} + \frac{Q(v)^2}{r^2} \right) dv^2 + 2dvdr + r^2 \left(d\vartheta^2 + \sin^2\vartheta d\varphi^2 \right),
\end{equation}

\noindent where $M(v)$ denotes the mass, and $Q(v)$ denotes the electric charge of the spacetime \cite{VB:1970bh}. We want to remark that the closest analogy to the axisymmetric Kerr-Vaidya spacetime presented in Section \ref{KV} is given by a constant charge, i.e. $Q'(v) = 0$. \\

Computing the Brown-York quasi-local mass in Vaidya-Bonnor spacetime, we follow the exact same procedure as for Kerr-Vaidya spacetime, see Appendix \ref{Proof_BY}. In this case, the closed 2-surfaces $S_r$ are just regular spheres, so that the Brown-York mass computes to \cite{Mathematica, xAct,xCoba,xPerm,xTensor}
\begin{equation}
 M_\textrm{BY} = M(v). 
\end{equation}

\noindent On the other hand, the Misner-Sharp-Hernandez (MSH) mass, see \cite{MS:1964mass,HM:1966mass,FGB:2021ql}, which is exclusively defined in spherical symmetry, computes as \cite{Mathematica, xAct,xCoba,xPerm,xTensor}
\begin{equation}
M_\textrm{MSH} := \frac{r}{2}\left( 1 - \nabla^a r \, \nabla_a r \right) = M(v) - \frac{Q(v)^2}{2r}.
\end{equation}

\noindent We observe that only in the asymptotically flat region, the MSH mass consistently agrees with the Brown-York mass $M(v)$. \\

Since Vaidya-Bonnor spacetime is spherically symmetric, there exists a Kodama symmetry generated by the vector field $k^a$, which is of the form
\begin{equation}
k^a = \left( \frac{\partial}{\partial v} \right)^a,
\end{equation}

\noindent i.e. the current $j^a = G^{ab}k_b$ is covariantly conserved. Following the same calculation as in Appendix \ref{ProofMassCharge}, we find that its corresponding charge $Q_k$ is equal to the MSH mass, i.e.
\begin{equation}
Q_k = \frac{1}{8\pi} \int_{\Sigma_r} j^a n_a \, d\textrm{vol}_{\Sigma_r} = M(v) - \frac{Q(v)^2}{2r} = M_\textrm{MSH}.
\end{equation}

\noindent Similarly, we can perform the calculations from Appendix \ref{ProofAngularMomentum} for the angular Killing vector field
\begin{equation}
\phi^a = \left( \frac{\partial}{\partial \varphi} \right)^a,
\end{equation}

\noindent in order to find
\begin{equation}
Q_\phi = 0.
\end{equation}

\noindent The physical interpretation of this is that Vaidya-Bonnor spacetime has zero angular momentum, as expected.

\section{Noether Charges in Kerr-Vaidya-de Sitter Spacetime}\label{IntegralsKVdS}
The exact expression for the charge $Q_{\tilde{K}}$ of the Kodama-like vector field $\tilde{K}^a$ in Kerr-Vaidya-de Sitter spacetime is, according to expression \eqref{ChargeIntegral}, given by \cite{Mathematica, xAct, xPerm, xTensor, xCoba}
\begin{equation}
Q_{\tilde{K}} = \frac{1}{8\pi} \int_{\Sigma_r} \frac{1}{\rho^6}\left( \left( \rho^4 + r^4 + r^2 a^2 \sin^2\vartheta - a^4\cos^2\vartheta \right) M'(\tilde{v}) - \frac{\rho^2 a^2 r Z \sin^2\vartheta}{\Theta} M''(\tilde{v}) \right) d\textrm{vol}_{\Sigma_r}.
\end{equation}

\noindent We observe that the first term, containing $M'(\tilde{v})$, is equal to the term containing $M'(v)$ in expression \eqref{ExplicitChargeIntegral} of Appendix \ref{ProofMassCharge}, up to the constant factor $Z^{-2}$ from the volume element. Hence, it follows directly that
\begin{equation}
Q_{\tilde{K}} = \frac{M(\tilde{v})}{Z^2} - \frac{1}{8\pi} \int_{\Sigma_r} \frac{a^2 r \sin^3\vartheta}{\rho^2 Z \Theta} M''(\underline{\tilde{v}}) \, d\underline{\tilde{v}}\,d\vartheta \,d\tilde{\psi}.
\end{equation}

\noindent The remaining integral can be computed by similar methods as in the previous Appendices \ref{ProofMassCharge} and \ref{ProofAngularMomentum}. We obtain
\begin{equation}
\label{deSitterMassCharge1}
Q_{\tilde{K}} = \frac{M(\tilde{v})}{Z^2} - M'(\tilde{v}) \left( \frac{Zr\arctan(Ha) - \Gamma H \arctan\left(\frac{a}{r}\right)}{2ZHa \, (r^2 H^2 - 1)} \right).
\end{equation}

\noindent While the second term appears to contain a singularity at $r=\frac{1}{H}$\footnote{Note that $r=\frac{1}{H}$ does \emph{not} correspond to the cosmological horizon $r_C$, since $\Delta_\Lambda\left(\frac{1}{H}\right)\neq 0$ \cite{Mathematica, xAct, xPerm, xTensor, xCoba}.}, it can be shown by L'Hospital's rule that the limit $r\rightarrow \frac{1}{H}$ of $Q_{\tilde{K}}$ is actually finite, given by
\begin{equation}
\lim_{r\rightarrow\frac{1}{H}} Q_{\tilde{K}} = \frac{M(\tilde{v})}{Z^2} - M'(\tilde{v}) \left( \frac{Ha - (1-H^2a^2) \arctan\left(Ha\right)}{4ZH^2 a} \right).
\end{equation}

\noindent Furthermore, we are able to verify by the same approach that in the limit $H\rightarrow 0$, the charge $Q_{\tilde{K}}$ consistently coincides with its counterpart from the asymptotically flat scenario, given in equation \eqref{KodamaCharge}, i.e. $\lim_{H\rightarrow 0} Q_{\tilde{K}} = Q_K$. \\

Regarding the angular momentum of Kerr-Vaidya-de Sitter spacetime, we obtain by a direct calculation \cite{Mathematica, xAct, xPerm, xTensor, xCoba} that the charge $Q_{\tilde{\phi}}$ of the current $\tilde{I}^a := \tilde{G}^{ab}\tilde{\phi}_b$, generated by the Killing vector field $\tilde{\phi}^a=\left(\frac{\partial}{\partial\tilde{\psi}}\right)^a$, is given by
\begin{align}
Q_{\tilde{\phi}} &= \frac{M(\tilde{v})a}{Z^2} - \frac{1}{8\pi} \int_{\Sigma_r} \frac{a^3 r \sin^5\vartheta}{\rho^2 Z \Theta} M''(\underline{\tilde{v}}) \, d\underline{\tilde{v}}\,d\vartheta \,d\tilde{\psi} \\
&= \frac{M(\tilde{v})a}{Z^2} - M'(\tilde{v})\left( \frac{H^3 a r^3 + r\left(Z^2 \arctan(Ha) - Ha \right) - \Gamma^2 H^3 \arctan\left(\frac{a}{r}\right)}{2 Z H^3 a^2 \left(H^2 r^2 - 1\right)} \right).
\end{align}

We find that $Q_{\tilde{\phi}}$ qualitatively behaves completely analogous to $Q_{\tilde{K}}$, i.e. the limit $r\rightarrow\frac{1}{H}$ of $Q_{\tilde{\phi}}$ is well-defined, and $Q_{\tilde{\phi}}$ converges to the charge $Q_\phi$ from the asymptotically flat case in the limit $H\rightarrow 0$ \cite{Mathematica, xAct, xPerm, xTensor, xCoba}.

{\footnotesize
\bibliographystyle{unsrt}
\bibliography{bibliography}}

\end{document}